\documentclass[prl,aps,floats,twocolumn]{revtex4}

\usepackage{graphicx}
\usepackage{amssymb} 
\usepackage{amsmath}

\begin{document}

\def\gtrsim{\mathrel{\hbox{\rlap{\hbox{\lower4pt\hbox{$\sim$}}}\hbox{$>$}}}}

\title{Cosmic Tides}
\author{
Ue-Li  Pen$^{1}$,
Ravi Sheth$^{2}$,
J. Harnois-D\'{e}raps$^3$,
Xuelei Chen$^4$ and 
Zhigang Li$^4$
}
\affiliation{
  $^1$Canadian Institute for Theoretical Astrophysics, Toronto, Canada\\
$^2$ICTP, Trieste, Italy \\
$^3$Physics department, University of Toronto, Canada \\
$^4$National Astronomy Observatories, Chinese Academy of Science, Beijing, China
}

\date{\today}

\begin{abstract}

We apply CMB lensing techniques to large scale structure and solve for
the 3-D cosmic tidal field.  We use small scale filamentary structures
to solve for the large scale tidal shear and gravitational potential.
By comparing this to the redshift space density field, one can measure
the gravitational growth factor on large scales without cosmic
variance.  This potentially enables accurate measurements of neutrino
masses and reconstruction of radial modes lost in 21 cm intensity
mapping, which are essential for CMB and other cross correlations.  We
relate the tidal fields to the squeezed limit bispectrum, and present
initial results from simulations and data from the SDSS.

\end{abstract}
\maketitle

\newcommand{\be}{\begin{eqnarray}}
\newcommand{\ee}{\end{eqnarray}}
\newcommand{\beq}{\begin{equation}}
\newcommand{\eeq}{\end{equation}}

{\it Introduction.} --
The large scale structure of the universe shows striking non-Gaussian
features, often described as the cosmic web.  The filamentary nature
arises from gravitational tidal shear.  It is the same principle that
leads to ocean tides on earth: the residual anisotropy of gravitational
forces in a free falling frame.  The strong non-Gaussian nature of this system has
traditionally led to a reduction in cosmological information that can
be extracted from a large
survey \cite{1999MNRAS.308.1179M,2008MNRAS.388.1819L}.  In this paper,
we turn the process around, and exploit the tidal non-Gaussianity to
improve the measurement of large scale structures.

The gravitationally induced displacement is a three component vector
field.  Only changes in displacement are observable, which are
described by the Jacobian of the displacement field.  The Jacobian can
result in a change of volume and of shape.  In this paper we shall
study the change of shape, which is more robust since many other
processes can lead to a change of number density.  This approach is
equivalent to the gravity wave shear computed in
\cite{2010PhRvL.105p1302M}.  We are using small scale structure
alignment to solve for the large scale tidal field.  This provides
many independent samples to accurately measure the large scale tidal
field.

Generally speaking, the shape tensor in three dimensions is described
by 5 numbers: three Euler angles and two axis ratios.  The
gravitational potential is a single number, which is five fold
overdetermined. The radial changes are affected by peculiar
velocities, which are beyond the scope of this paper, hence we only
use the shear in the plane of the sky.  This tangential shear is
described by two numbers, and we will use the gravitational lensing
notation, where the two variables are called $\gamma_1,
\gamma_2$\cite{1992grle.book.....S}.  A linear transformation
decomposes them into a divergence, or E-like mode (called $\kappa$),
and a curl, or B-like
mode\cite{1991ApJ...380....1M,1998MNRAS.301.1064K,2002ApJ...567...31P}.

{\it Tidal Reconstruction.} -- The reconstruction of tidal shear is
described by the same formulation as the reconstruction of
gravitational lensing induced shear.  It leads to a local anisotropy
of quadratic statistics.  As shown in \cite{2010PhRvD..81l3015L},
there exists an optimal quadratic estimator that is solvable but
distinct for both Gaussian and Non-Gaussian fields.  In the first
case, the optimal estimator can be expressed in terms of the power
spectrum of the density field alone.  In general, however, the optimal
weights need to be computed from simulations
\cite{2011arXiv1109.5746H}.  

\begin{figure}
\begin{center}
    \includegraphics[width=3.1in]{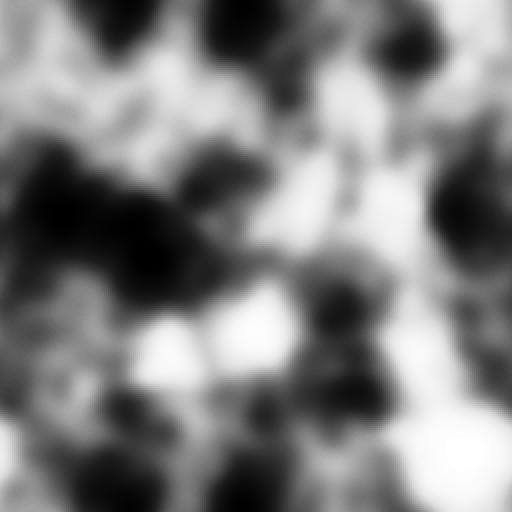}
    \includegraphics[width=3.1in]{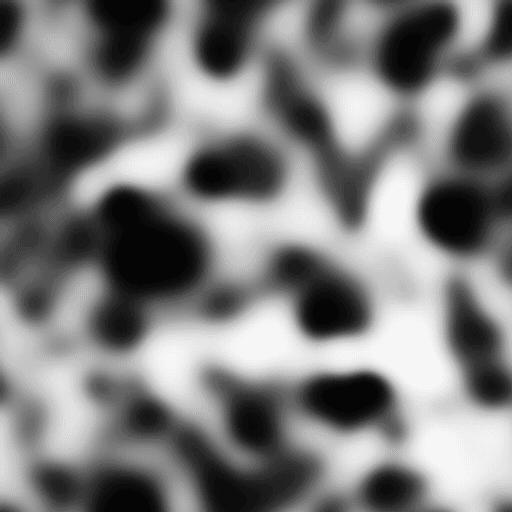}
  \end{center}
  \caption{
Density field slice smoothed on 8$h^{-1}$ Mpc in a 322$h^{-1}$ Mpc simulation
box.  Top panel:  Original dark  
matter field. Bottom panel: reconstructed smoothed field. }
\label{fig:rhok}
\end{figure}

{\it Heuristic kernel.} --
Here we work through a simple, slightly sub-optimal, scenario.  We
start with the density field $\delta(x)$.  In analogy to 
CMB lensing, quadratic estimators are outer products of gradients.
The first step is a convolution, which filters for the small scale
structure (the gradient downweights the large scales).   The large
scale gravitational field $\phi$ will be a linear convolution of this
local small scale quadratic estimator. For simplicity,
we use a Gaussian window  
$W(\vec{r})\equiv \exp\left(\frac{-|r|^2}{2\sigma^2}\right)$.
We obtain a smoothed density field:
$
{\bar{\delta}}(x)\equiv\int {W}(x-x') \delta(x')d^3x'.
$

Because the quadratic estimator heavily weights the high density
regions, we further apply a Gaussianization technique
\cite{1992MNRAS.254..315W} to the non-Gaussian (smoothed) density
field by taking $\delta_g\equiv\log (1+\bar{\delta})$.
Following gravitational lensing procedures, we construct the two shear
components 
as $\gamma_1\equiv(\partial_x\delta_g)^2-(\partial_y \delta_g)^2$ and
$\gamma_2\equiv2(\partial_x\delta_g)\partial_y\delta_g$.  We can then
reconstruct the dark matter field $\kappa$, in analogy with the
convergence field, which is a linear convolution of these two
estimators. 
In
terms of differential operators, we define
$d\equiv(\partial_x^2-\partial_y^2)\gamma_1+2\partial_x\partial_y\gamma_2$,
and solve Poisson's equation for the density
$(\partial_x^2+\partial_y^2)\kappa_z=N d$ on each $z$ slice.  The 3-D
dark matter density field $\kappa(x,y,z)$ is given by one more ratio
of wave numbers
$(\partial_x^2+\partial_y^2)\kappa=(\partial_x^2+\partial_y^2+\partial_z^2)\kappa_z$. $N$ 
is a normalization constant which is derived in
\citet{2008MNRAS.388.1819L}.

{\it Simulations.} -- We ran N-body simulations with the {\small
  CUBEP3M} code\cite{2005NewA...10..393M}, evolving $256^3$ particles
on a $512^3$ grid in a 322$h^{-1}$ Mpc box.  The smoothing window $W$
has a width $\sigma=1.25h^{-1}$Mpc.  Figure \ref{fig:rhok} shows a
slice of the original and reconstructed smoothed density fields, both
once again smoothed on a 8$h^{-1}$ window to reduce the small scale
noise.  We found little dependence on the windowing 
scale until we smooth on linear scales, which is related to the
information saturation phenomenon discussed below.

\begin{figure}
\begin{center}
    \includegraphics[width=3.1in]{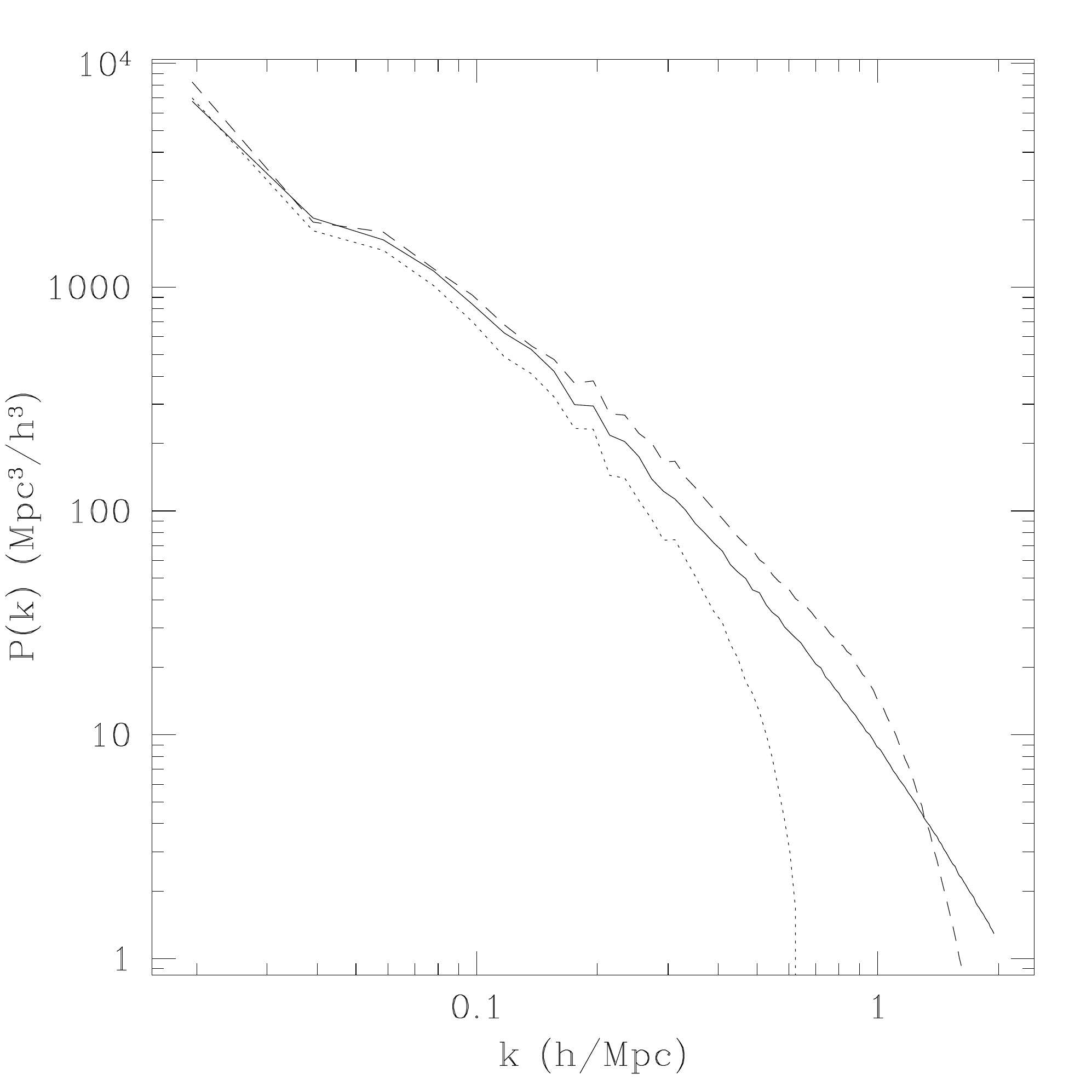}
  \end{center}
  \caption{Simulation results: Dark matter power spectrum (solid line), tidal
    reconstructed power spectrum (dashed line), cross power spectrum
    (dotted line).  We see good agreement between the reconstructed power
    spectrum and the dark matter power.
}
\label{fig:pk}
\end{figure}

Figure \ref{fig:pk} shows the raw and reconstructed power spectra, as
well as the cross correlation spectrum.  The amplitude of the
reconstructed spectrum was scaled to match that of the dark matter.  In
principle, this amplitude is determined for truly Gaussian random fields, but in
practice these assumptions do not hold precisely.
The amplitude of the cross correlation is now completely determined,
and we see a good cross correlation over a decade in wave number.

\begin{figure}
\begin{center}
    \includegraphics[width=3.1in]{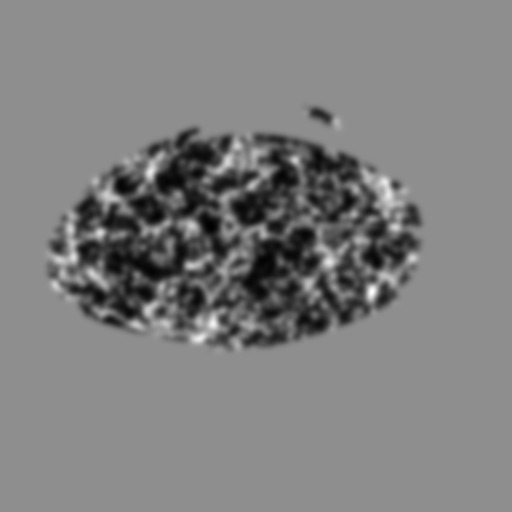}
    \includegraphics[width=3.1in]{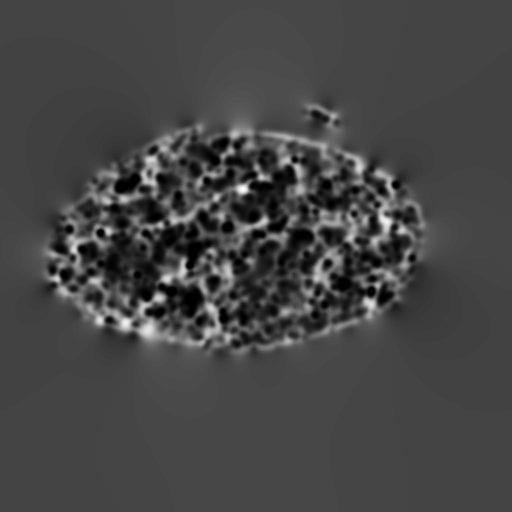}
  \end{center}
  \caption{
$z=0.1$ slices of SDSS survey embedded in a 900$h^{-1}$ Mpc box,  smoothed by a window function. Top panel:  Original galaxy
field. Bottom panel: dark matter map reconstructed from small scale
galaxy density field alignments.
}
\label{fig:kappasdss}
\end{figure}

{\it Data.} --
We also applied the technique to a volume limited subsample of the
Sloan Digital Sky Survey (SDSS) in which 112043 galaxies were selected.  We
constructed a Cartesian mapping as follows: A 0.9$h^{-1}$ Gpc box is
centered on the 
north galactic pole, at $z=0.1$.  We used the redshift range
$0.7<z<1.3$.  The angular position is mapped at 300$h^{-1}$ Mpc per radian,
and the radial position at 0.01$h^{-1}$ Mpc/(km/sec).  A random catalog with five
times as many galaxies was used as a density normalization.  The log
of the smoothed random density is subtracted from the log of the
smoothed galaxy count.  Pixels with less than half the peak random
galaxy density count were masked. Figure \ref{fig:kappasdss} shows a slice of the galaxy and
reconstructed density fields at $z=0.1$.

\begin{figure}
\begin{center}
    \includegraphics[width=3.1in]{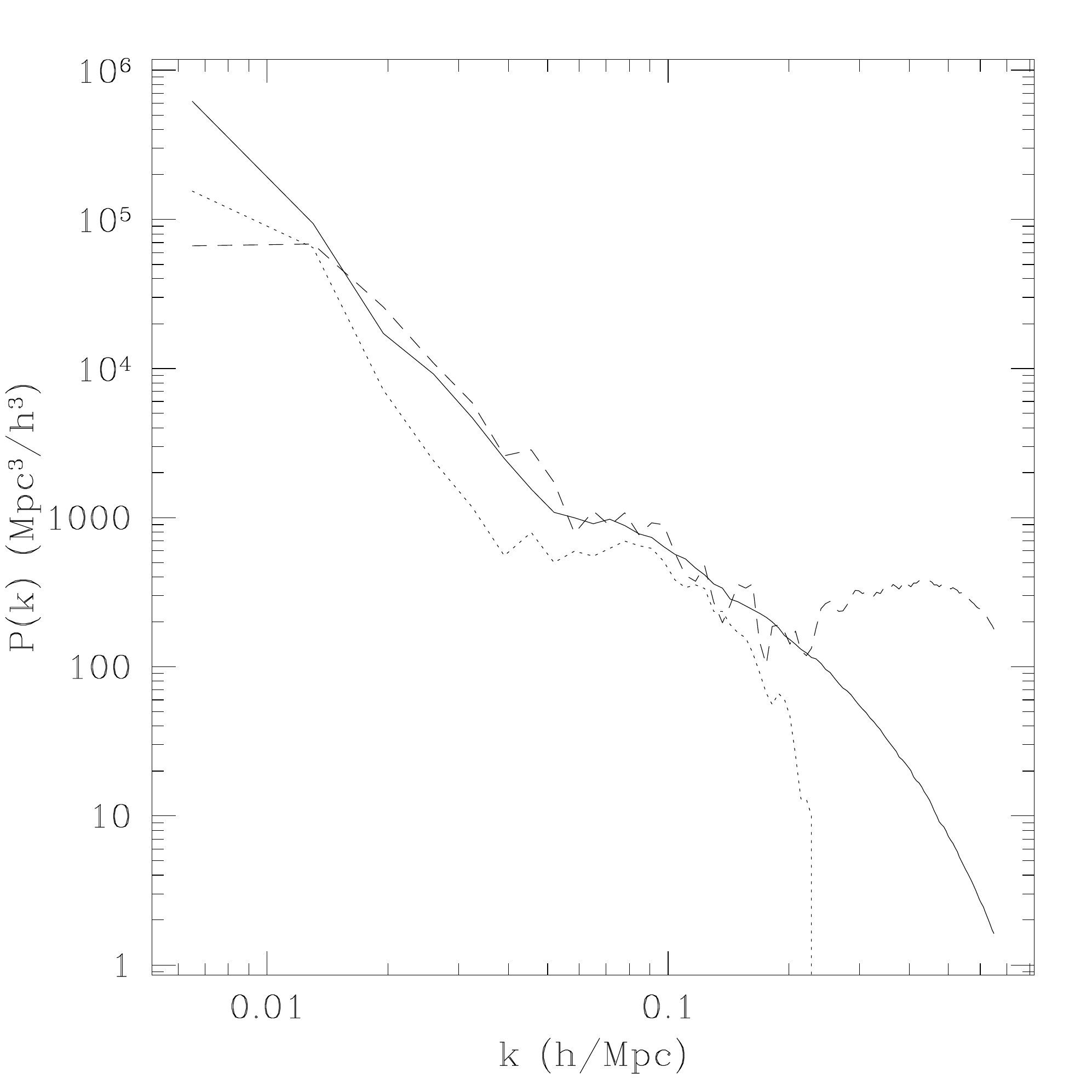}
  \end{center}
  \caption{SDSS results: galaxy power spectrum (solid line), tidal
    reconstructed power spectrum (dashed line), cross power spectrum
    (dotted line).  We see good agreement for the reconstructed power
    spectrum. 
}
\label{fig:pksdss}
\end{figure}

The original and reconstructed power spectra are shown in Figure \ref{fig:pksdss}, where  we
have again scaled the reconstructed power spectrum.  When comparing
the biased galaxies, the bias will always be a free parameter, such that the
non-linear effects of normalization can be absorbed as a relative
bias.  The cross power spectrum has a cross correlation coefficient
$r>0.5$ for all $k$ up to 0.2$h$/Mpc, thereby demomstrating a
qualitative successful dark matter reconstruction.

{\it Discussion.} --
We have seen a successful application of a heuristic tidal field
estimator.   Here we will discuss the theoretical framework, and
future directions.

The large scale tidal shear field is the expectation value of the
quadratic outer product 
\be
T_{ij}\propto\langle(\partial_i\bar{\delta})\partial_j \bar{\delta}\rangle.
\ee
Redshift space
distortions complicate every components that include $i=3$.
Our shear components are related as $\gamma_1=T_{11}-T_{22}$ and
$\gamma_2=2T_{12}$.  
It was shown that this two index tensor is proportional to the
traceless tidal shear tensor (see eqn(35)
of\cite{2008MNRAS.388.1819L}) derived from the gravitational potential:
$
\tilde{T}_{ij}=\partial_i \partial_j \phi-\frac{\delta_{ij}}{2}\nabla^2 \phi.
$
The trace, which corresponds to the local mean density, in principle transforms in 
a similar way, however we will defer such study to the future.
The trace free estimator is quadratic in the density field, and does not
require second order perturbation theory to compute.  Being trace
free, it does not backreact on the mean density.  
We are only considering waves in $\phi$ on much larger scales $k_\phi$
than the galaxy density $\delta$: $k_\delta \gg k_\phi$.  

The reconstructed density field is derived from a convolution:
\be
\kappa = \int T_{ij}(x') K_{ij} (x-x') d^3 x'
\ee
where the kernel $K_{ij}$ is given in Fourier space as
\be
K_{ij}(k)=\frac{k^2}{k_x^2+k_y^2}\left(\hat{k}_i\hat{k}_j-\frac{\delta_{ij}}{2}\right).
\ee

In the limit that the tidal mode is linear and has a wave length much
longer then the smoothing scale of the Gaussian window, the
perturbative coupling arises solely from the induced gravitational
displacement.  

Free falling observers are only sensitive to tidal forces, and the
tidal force on the window scale is just a stretching of coordinates.
The tidal distortion field can be viewed as an integral over the
history of gravity.  We note that the actual density field only
follows linear evolution weakly, even on apparently linear
scales\cite{2004MNRAS.355..129S}.

{\it Relation to CMB Lensing.} -- The effect described here is
identical to the CMB lensing phenomenon, which also induces a three
point correlation between two points on the CMB and one in the
intervening space\cite{2006PhR...429....1L}.  In our case, two wave
vectors are on small scales, and one on large scales, all in the same
volume.

{\it Connection to Non-Gaussianity.} -- Our separation of scales is
equivalent to bispectrum calculations in the squeezed limit.  We are
measuring the variance modulated by a large scale, small $k$ mode,
again in analogy with the CMB case\cite{2011JCAP...11..025C}.  The
coordinate change is the leading order effect relevant for the shear,
whose measurement corresponds to using only the term dependent on the
angle $\alpha$ between the small $k_\phi$ vector and the two large
ones $k_\delta$.  Figure \ref{fig:pksdss} is the covariance of
$\langle \delta\rangle$, and $\kappa$ is in turn a quadratic function
of $\delta$ at large wave numbers.  We have effectively integrated the
bispectrum over its quadrupole $\int \langle \delta(k_\phi)
|\delta(k_\delta)|^2 \rangle \cos(2\alpha) d\alpha$.

This is the leading effect contributing to the
information saturation phenomenon\cite{1999MNRAS.308.1179M}.  It was
found that the variance in the power spectrum on quasi-linear scales
was up to three orders of magnitude larger than expected for Gaussian
random fields.  This was also observered for variances in the shear
estimators\cite{2008MNRAS.388.1819L,2011arXiv1109.5746H}.  It had been
proposed that this saturation was due to Poisson noise in the virialized
halos\cite{2006MNRAS.370L..66N}, however  Poisson noise does not lead to
an increase in the shear variance, hence can not be the complete picture.
The shear has been numerically observed to saturate at a similar
information content as the mean variance.  Furthermore, reconstruction
techniques are able to increase the information
content\cite{2012MNRAS.419.2949N}, which is not expected in a Poisson
model.

The cosmic tides picture can qualitatively explain this information
saturation effect.  A 10\% tidal distortion results in a systematic change
of the variance by $\sim 10\%$.  When more than 100 modes are
measured, their variance is dominated by the larger scale shear.  An
analogous effect applies to the power spectrum variances.  This
implies that saturation is dominated by modes near the peak of the
power spectrum, $k\sim 0.02$.

{\it Recovering Lost 21cm Modes.} --
21cm intensity mapping has emerged as a promising technique to map the
large scale structure of the universe, at redshifts $z$ from 1 to 10.
Unfortunately, many of the key cross correlations with photo-x galaxy and the
CMB have been thought to be impossible due to foreground contamination for
radial modes with small wave
numbers\cite{2009ApJ...690..252L,2008MNRAS.384..291A}.  Our proposed tidal 
reconstruction technique works best in this regime, and opens up a new set of
possibilities. 

{\it Reducing Sample Variance.} --
The high cross correlation seen in the simulation suggests that one
can measure the dark matter density on large scales without redshift
space distortions.  By comparing the galaxy power spectrum to the
$\kappa$ power spectrum as a function of angle to the line of sight,
one can solve for the velocity contribution.   This allows for a
measurement of velocity on the same mode as the density, hence
determines the velocity growth factor without sample variance,
analogous to \citet{2009JCAP...10..007M}.
In principle, this could enable precision measurements of neutrino
masses. 

{\it Conclusion.} --
We have presented a new framework to measure the effects of gravity
through its tidal effect on the structure of galaxy clustering.  We
isolated this effect theoretically from other complications by
projecting out mean density changes.  The effect becomes analogous to
the lensing shear on large scale structure or the CMB.  We have tested
the dark matter density reconstruction in simulations and with the SDSS
data.  We found good agreement, with reconstruction noise less than
sample variance.  This opens up the window for precision measurements
of the transfer function, potentially measuring neutrino masses or
testing modified gravity.

ULP thanks NSERC, SHAO and NAOC for financial support.  XC was
supported by grants NSFC 11073024, CAS KJCX3-SYW-N2.  N-body
computations were performed on the TCS supercomputer at the SciNet HPC
Consortium. SciNet is funded by: the Canada Foundation for Innovation
under the auspices of Compute Canada; the Government of Ontario;
Ontario Research Fund - Research Excellence; and the University of
Toronto.  We thank Kiyoshi Masui for helpful comments.

\newcommand{\araa}{ARA\&A}   
\newcommand{\afz}{Afz}       
\newcommand{\aj}{AJ}         
\newcommand{\azh}{AZh}       
\newcommand{\aaa}{A\&A}      
\newcommand{\aas}{A\&AS}     
\newcommand{\aar}{A\&AR}     
\newcommand{\apjs}{ApJS}     
\newcommand{\apjl}{ApJ}      
\newcommand{\apss}{Ap\&SS}   
\newcommand{\baas}{BAAS}     
\newcommand{\jaa}{JA\&A}     
\newcommand\jcap{{J. Cosmology Astropart. Phys.}}%
\newcommand{\mnras}{MNRAS}   
\newcommand{\na}{New Astronomy} 
\newcommand{\pasj}{PASJ}     
\newcommand{\pasp}{PASP}     
\newcommand{\paspc}{PASPC}   
\newcommand{\qjras}{QJRAS}   
\newcommand{\sci}{Sci}       
\newcommand{\sova}{SvA}      
\newcommand{\aap}{A\&A}
\newcommand{\physrep}{Phys. Rep.}

\bibliography{tides}

\label{lastpage}

\end{document}